\documentstyle[12pt,emulateapj,psfig]{article}

\newcommand\pp     {$\pm$}
\newcommand\pers     {s$^{-1}$}
\newcommand\micros  {$\mu$s}

\righthead{Rapid X-ray variability of V4641 Sagittarii }
\slugcomment{Resubmitted to ApJ Letters, 1999 November 5}

\begin{document}

\title{The rapid X-ray variability of V4641 Sagittarii (SAX
J1819.3-2525 = XTE J1819-254)}

\author{Rudy Wijnands\footnote{Chandra Fellow}}

\affil{Center for Space Research, Massachusetts Institute of
Technology, 77 Massachusetts Avenue, Cambridge, MA 02139-4307, USA;
rudy@space.mit.edu}

\author{Michiel van der Klis}

\affil{Astronomical Institute ``Anton Pannekoek'', University of
Amsterdam, and Center for High Energy Astrophysics, Kruislaan 403,
NL-1098 SJ Amsterdam, The Netherlands; michiel@astro.uva.nl}

\begin{abstract}
We report on the rapid X-ray variability of the variable star and
X-ray transient V4641 Sagittarii (SAX J1819.3--2525; XTE J1819--254)
as observed on 15 Sept. 1999 by the proportional counter array (PCA)
on board the {\it Rossi X-ray Timing Explorer} ({\it RXTE}). During
the first $\sim$900 s of the first PCA observation, V4641 Sgr showed
very strong X-ray fluctuations by a factor of 4 on time scales of
seconds to about 500 on time scales of minutes. The spectrum of the
source during this flaring episode became harder when the count rate
decreased. After this flaring episode, V4641 Sgr entered a quiescent
state in which it remained for the rest of this, and the subsequent
PCA observations. The X-ray spectrum was considerably softer in this
quiescent state than during the flaring episode. The intrinsic X-ray
luminosity (both during the flaring episode and the quiescent state)
and the rapid X-ray variability do not strongly constrain the nature
of the compact object (neutron star or black hole) in the system,
although a black hole seems to be more likely.  The very short
duration of the bright X-ray phase of V4641 Sgr and its likely close
distance suggest that many similar objects could be present in our
galaxy, most of which are not noticed when they are in X-ray outburst
due to the short duration of these outbursts.  A considerable number
of the black holes present in our galaxy might be contained in systems
similar to V4641 Sgr.

\end{abstract}


\section{Introduction \label{intro}}

In Feb. 1999, a new X-ray transient was independently discovered by
{\it BeppoSAX} (SAX J1819.3--2525; in 't Zand et al. 1999) and {\it
RXTE} (XTE J1819--254; Markwardt, Swank, \& Marshall 1999a). Its
position is consistent with that of the variable star V4641 Sgr (note
that V4641 Sgr had been misidentified as GM Sgr [see IAU Circular
7277]). Its 2--10 keV flux varied between $<$1 and 80 mCrab (in 't
Zand et al. 1999; Markwardt et al. 1999a).  On 15 Sept.  1999, the
source rapidly increased in the optical (Stubbings 1999) and with the
{\it RXTE} All-Sky Monitor (ASM) its 2--12 keV intensity was observed
to increase very rapidly (within 7 hours) from 1.6 to 12.2 Crab
(Smith, Levine, \& Morgan 1999a,b). Subsequent ASM measurements showed
that within two hours of this flare, the flux had declined down to a
level only marginally detectable with the ASM ($<$50 mCrab; Smith et
al. 1999c).

Optical and infrared spectra taken during this bright X-ray event show
emission lines (Ayani \& Peiris 1999; Liller 1999; Djorgovski et
al. 1999; Charles, Shahbaz, \& Geballe 1999), reminiscent of accretion
of matter onto a compact object, demonstrating that V4641 Sgr is
indeed the optical counter part.  On 16 Sept. 1999, VLA observations
were taken and a radio source was discovered (Hjellming, Rupen, \&
Mioduszewski 1999a) at a position consistent with V4641 Sgr. Follow up
VLA and ATCA observations showed that its flux was rapidly declining
on time scales of hours to days (Gaensler et al. 1999; Hjellming,
Rupen, \& Mioduszewski 1999b; Hjellming et al. 1999c).  The VLA
observations also showed that it was resolved, demonstrating the
presence of ejecta (Hjellming et al. 1999b,c).

On 15 Sept. 1999, a short (3000 s) pointing was taken with the {\it
RXTE} proportional counter array (PCA). A rapidly variable source was
observed with a flaring and a quiescent episode (Markwardt, Swank, \&
Morgan 1999b). No pulsations or quasi-periodic oscillations (QPOs)
were detected but red noise below 30 Hz was present.  Here, we discuss
in more detail the rapid X-ray variability as observed with the PCA
during this observation.

\section{Observation and Results \label{observations}}

Several public TOO PCA observations were scheduled between 15 and 18
Sept. 1999 for a total of $\sim$33 ksec on-source time. However, only
the first 1500 s of the first observation (on 15 Sept. 1999
21:18--22.43 UTC; see also Markwardt et al. 1999b) showed very high
fluxes and strong variability.  We limited our analysis to these first
1500 s in order to study the X-ray variability of the source. During
the second 1500 s of this observation and during the later
observations, V4641 Sgr could still be detected with the PCA, but at a
very low flux level ($<$100 counts \pers~for 5 detectors on; 2--60
keV) and no significant variability could be detected (upper limits of
5\%--40\% rms [2--22.1 keV; 0.01--100 Hz; depending on total on-source
time and source count rate during the individual observations] on band
limited noise similar to that detected during the first $\sim$900 s of
the first observation [see below]).

During the PCA observations, data were accumulated in several
different modes which were simultaneously active. Here, we used data
obtained with the 'Binned' modes B\_250US\_1M\_0\_249\_H (one photon
energy channel [effective range of 2--60 keV] and 244 \micros~time
resolution) and B\_4M\_8B\_0\_49\_H (eight channels covering 2--22.1
keV and 3.9 ms~time resolution).  These data were used to calculate
128 s FFTs to create the power spectra and the cross spectra, the 3.9
ms data were used to create light curves, a hardness-intensity
diagram, and a hardness curve.  The power spectra were fitted with a
function consisting of a constant (the dead-time modified Poisson
level) and a broken power law (the band-limited noise below $\sim$100
Hz). A Lorentzian was used to determine upper limits (95\% confidence
level) on the rms amplitude of QPOs above 100 Hz, assuming a QPO width
of 50 Hz. To correct for the small dead-time effects on the lags, we
subtracted the average 50--125 Hz cross-vector from the cross spectra
(van der Klis et al. 1987).

\begin{figure}[]
\begin{center}
\begin{tabular}{c}
\psfig{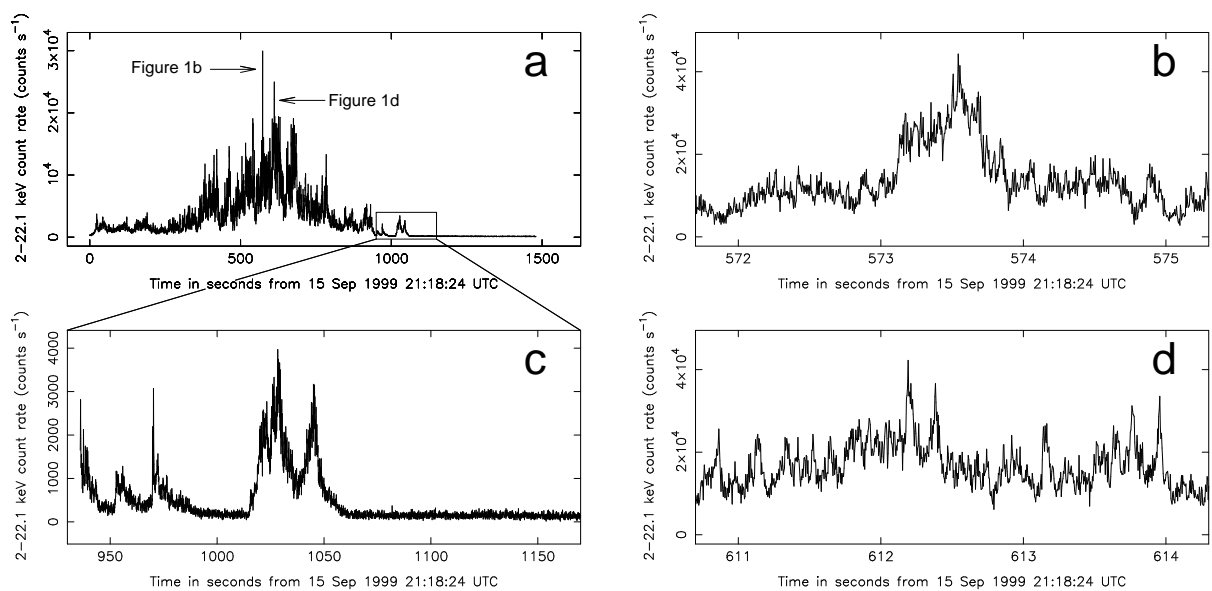}
\end{tabular}
\figcaption{The 2--22.1 keV light curve of V4641 Sgr obtained from the
3.9 ms Binned mode data ({\it a}). Blow ups of the two brightest
flares are shown in {\it b} and {\it d} and a blow up of the last
flare and the quiescent emission is show in {\it c}. The time
resolution is 0.25 seconds in {\it a}, 1/256 seconds (3.9 ms) in {\it
b} and {\it d}, and 0.0625 seconds in {\it c}. The count rates are for
5 detectors but are not corrected for background or dead-time. The
background varied between 50 and 65 counts \pers~(2--22.1 keV) during
the observation. The dead-time correction varied between 1\% and 10\%.
\label{fig:lc} }
\end{center}
\end{figure}

The 2--22.1 keV light curve of the first 1500 s of the first PCA
observation is shown in Figure~\ref{fig:lc}{\it a}, showing the
flaring behavior of V4641 Sgr between 300 and 900 s from the start.
After 1000 s, the source enters a quiescent state, with one last flare
between 1010 and 1060 s (Fig.\ref{fig:lc}{\it c}). After that (and
also in the second 1500 s of this observations, and in the other
observations) it remained in this state at very low count rates (see
also Markwardt et al. 1999b).  During the flaring episode, it varied
rapidly in count rate. On time scales of 5--10 minutes it varied in
luminosity by a factor of up to 500. Within one second, it sometimes
increased and then decreased by more than a factor of 4
(Figs.~\ref{fig:lc}{\it b} and {\it d}).

\begin{figure}[]
\begin{center}
\begin{tabular}{c}
\psfig{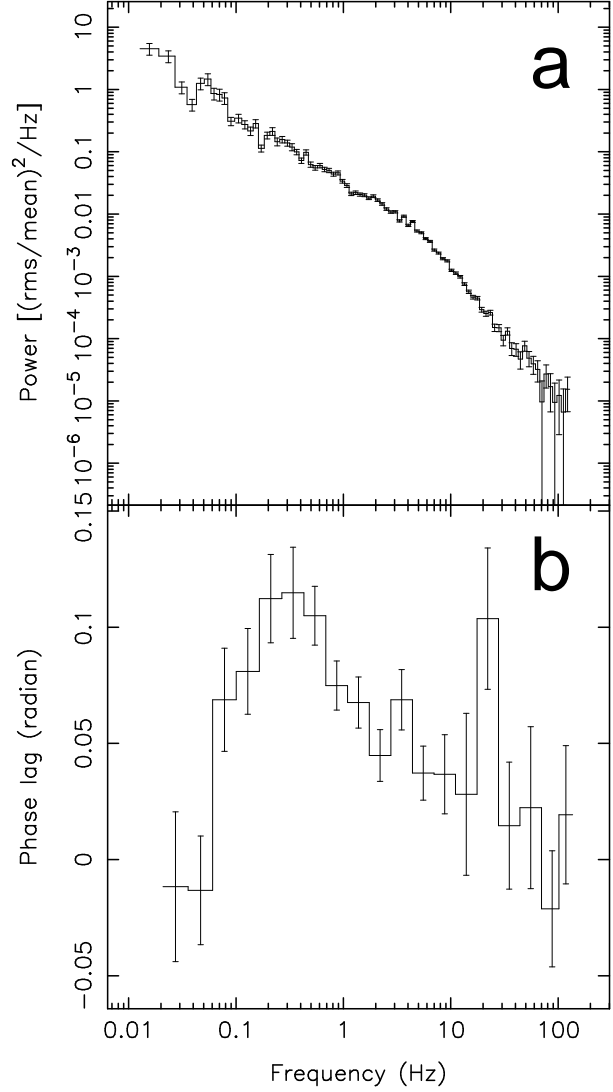}
\end{tabular}
\figcaption{The power spectrum ({\it a}) and the cross spectrum ({\it
b}) of the first 896 seconds of the first observation of V4641 Sgr
(obtained from the 3.9 ms Binned mode data). The power spectrum was
calculated for the energy range 2--22.1 keV. The cross spectrum was
calculated between the energy bands 2--9.7 keV and 9.7--22.1
keV. Positive phase lags mean that the hard photons lag the soft ones.
\label{fig:powercrossspectra} }
\end{center}
\end{figure}

\begin{figure}[]
\begin{center}
\begin{tabular}{c}
\psfig{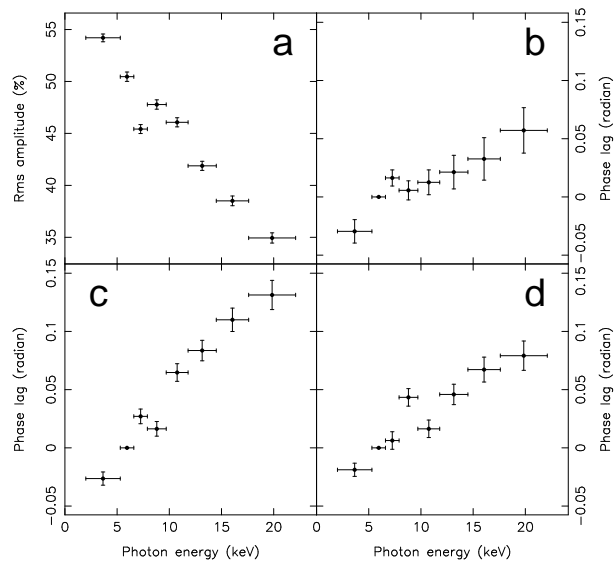}
\end{tabular}
\figcaption{The rms amplitude ({\it a}) and the phase lags between
0.02--0.1 Hz ({\it b}), 0.1--1.0 Hz ({\it c}), and 1.0--10.0 Hz ({\it
d}) of the broken power law as a function of photon energy. As a
reference band in {\it b}--{\it d}, we used the energy band 5.3--6.6
keV.  Positive phase lags mean that the hard photons lag the soft
ones.
\label{fig:rmsenergy} }
\end{center}
\end{figure}

The strong variability is also evident from the power spectra obtained
from the first 896 s (see Fig.~\ref{fig:powercrossspectra}{\it
a}). Strong band-limited noise (47.2\%\pp0.8\% rms amplitude;
0.01--100 Hz; 2--22.1 keV) can clearly be seen. Its shape fits a
broken power law with the break at 5.1\pp0.2 Hz and an index below and
above the break of 1.03\pp0.02 and 2.16\pp0.03, respectively. The rms
amplitude decreases from $\sim$54\% to $\sim$35\% as a function of
energy, increasing from $\sim$4 keV to $\sim$20 keV
(Fig.~\ref{fig:rmsenergy}{\it a}). By subtracting the 3.9 ms data
(2--22.1 keV) from the 244 \micros~data (2--60 keV), we could make a
power spectrum for the data above 22.1 keV. The rms amplitude of the
noise was even less, 30.6\%\pp0.5\%, above 22.1 keV (effective range
of 22.1--60 keV).  To show the decrease of amplitude below 20 keV more
clearly, we excluded this point from the figure.  The noise also has
significant hard phase lags between photons with energies of 2--9.7
keV and those with energies of 9.7--22.1 keV
(Fig.~\ref{fig:powercrossspectra}{\it b}). The lags increased from
being consistent with zero below 0.1 Hz to $\sim$0.1 rad at $\sim$0.03
Hz. Above this frequency, the lags decreased again and above about 10
Hz the lags were consistent with zero again. We studied the energy
dependence of the lags for four frequencies intervals: 0.02--0.1 Hz
(Fig.~\ref{fig:rmsenergy}{\it b}), 0.1--1.0 Hz
(Fig.~\ref{fig:rmsenergy}{\it b}), 1.0--10.0 Hz
(Fig.~\ref{fig:rmsenergy}{\it b}), and 10.0--50.0 Hz (not shown, the
lags were always consistent with zero).  Although the lags in the
range 0.02--0.1 Hz were barely significant
(Fig.~\ref{fig:rmsenergy}{\it b}) a clear trend with energy was
present.

\begin{figure}[]
\begin{center}
\begin{tabular}{c}
\psfig{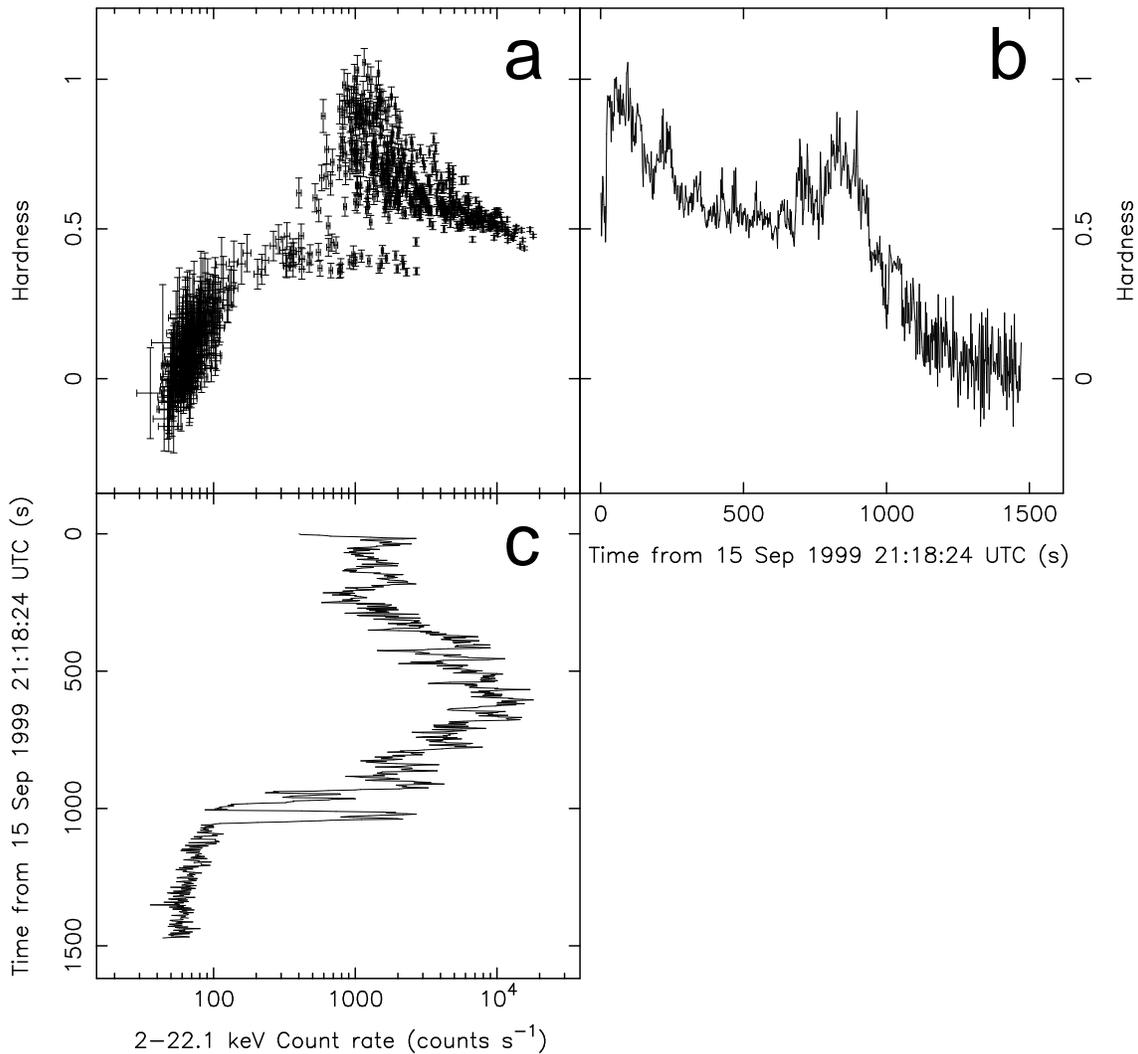}
\end{tabular}
\figcaption{Hardness-intensity diagram ({\it a}), the hardness curve
({\it b}), and the light curve ({\it c}) as obtained with the 3.9 ms
Binned mode data. The hardness is the count rate ratio between
9.7--22.1 keV and 2--9.7 keV. The intensity is the count rate for the
energy range 2--22.1 keV. The time resolution is 2 seconds. The count
rates are background subtracted but not dead-time corrected. The
dead-time correction is between 1\% and 10\%.
\label{fig:hid}}
\end{center}
\end{figure}

In the power spectrum obtained from the 244 \micros~data, no QPOs
above 100 Hz were detected (upper limits of 2\% rms). However, these
data cover the whole {\it RXTE} energy band.  Therefore, such QPOs
could have been present but undetectable if their strength depended
strongly on energy, similar to the kHz QPOs observed in many neutron
star and some black hole systems (van der Klis 1998, 1999; Remillard
et al. 1999; Homan, Wijnands, \& van der Klis 1999).  From the 3.9 ms
data, we made a hardness-intensity diagram (Fig.~\ref{fig:hid}{\it a})
and a hardness curve ({\it b}) with a time resolution of 2 s.  For
comparison, we also plotted the 2--22.1 keV light curve
(Fig.~\ref{fig:hid}{\it c}).  As hardness we used the count rate ratio
between 9.7--22.1 keV and 2--9.7 keV and as intensity the 2--22.1 keV
count rate, but plotted on a logarithmic scale to show the variations
more clearly. From Figure~\ref{fig:hid}{\it a}, it is clear that
during the flaring episode the hardness decreased when the count rate
increased. When the transition to the quiescent state occurred the
source became softer. During the last flare (in the quiescent
state; Fig.~\ref{fig:lc}{\it c}), the overall hardness was less than
for the flares during the flaring episode.

\section{Discussion \label{discussion}}

We have presented the X-ray variability of V4641 Sgr during its 1999
Sept. 15 bright event.  It shows strong variations by a factor of 4 on
time scales of seconds to variations by a factor of $\sim$500 on time
scales of minutes. These strong variations are also evident in the
power spectrum by the presence of strong (30\%--55\% rms amplitude,
depending on energy) band-limited noise.  Assuming the most likely
distance of 0.5--1.0 kpc (Hjellming et al. 1999c; obtained by
performing an HI absorption experiment against the radio counter part)
and using the maximum flux during our PCA observation (Markwardt et
al. 1999b), the maximum intrinsic luminosity (2--60 keV) was 0.3--1.0
$\times 10^{37}$ erg \pers~(note that it was at considerable higher
luminosities during the ASM measurements).  This luminosity, the
strong variability on short time scales, and the presence of
optical and infrared emission lines (see \S~\ref{intro}) suggest that
the X-rays are produced by accretion onto a compact object.

The exact nature of the compact object is difficult to determine. The
likely intrinsic luminosity and the hardness of the spectrum
(Markwardt et al. 1999b; see Smith et al. 1999c for a more detailed
analysis) strongly indicate that accretion onto a white dwarf, a nova
explosion, and thermonuclear burning on a white-dwarf surface cannot
account for the X-ray emission and its properties.  Compared to the
intrinsically brightest low magnetic field neutron star and black hole
systems, the luminosity of V4641 Sgr during our PCA observations was
relatively low (the brightest systems have an luminosities of $>$
$10^{38-39}$ erg \pers), which suggests that also the accretion rate
was relatively low.  At such low accretion rates, the rapid
variability (van der Klis 1995; Wijnands \& van der Klis 1999),
including the phase lags (Ford et al. 1999), and the spectrum (e.g.,
Barret \& Vedrenne 1994) of the low magnetic field neutron star and
the black hole systems are very similar (van der Klis 1994), making it
difficult to determine the exact nature of the compact object in V4641
Sgr.  Although the very strong variability suggests a black hole
primary in this system (usually such strong variability is only
observed for the black-hole systems), it cannot be excluded that some
neutron star systems can also exhibit such strong variability. The
decrease of the strength of the variability in V4641 Sgr with energy
is similar to what has been observed in several black-hole systems in
their low state (e.g., Nowak et al. 1999a; Nowak, Wilms, \& Dove
1999b), but the neutron star systems have not been studied in enough
detail in this respect to allow detailed comparisons between the
different types of systems.
		
It was suggested that the nature of the compact objects in X-ray
transients could be determined by studying and comparing the X-ray
emission properties of those systems in their quiescent state (see,
e.g., Rutledge et al. 1999). The luminosity for V4641 Sgr in its
quiescent episode during the first PCA observation was 0.5--2.2
$\times 10^{34}$ erg \pers~but monitoring observations with the PCA of
the galactic-center region showed that during the last 7 months the
luminosity had dropped occasionally to 0.6--2.4 $\times 10^{33}$ erg
\pers~ (see Markwardt et al. 1999b for the fluxes used).  However,
these luminosities do not give more insight in the nature of the
compact object. Both the observed or derived upper limits on the
luminosities for neutron star and black hole X-ray transients in
quiescence are consistent with the values detected for V4641 Sgr in
its quiescent episode.  More detailed studies in quiescence are needed
to determine the lowest observed luminosity in quiescence. If V4641
Sgr contains a black hole then it is expected that the luminosity
should drop sometimes significantly below 10$^{32}$ erg \pers, as
observed in other black-hole transients in quiescence.

Thus, the properties of V4641 Sgr as observed with the PCA are very
similar to those observed for an accreting compact object which
accretes matter from its companion star at a low rate.  Although a
neutron star primary cannot be excluded, the strong variability and
the low intrinsic luminosity make an interpretation of V4641 Sgr as a
black-hole candidate in the low state most favorable.  The most
promising way to determine the exact nature is to dynamically (during
the quiescent state; from the optical line spectrum of the companion
star) constrain the mass of the primary. If the primary is truly a
black hole, then several features have been observed for V4641 Sgr
which are not commonly observed in black hole systems in their low
state. For example, its very strong variability have only been
observed in a few other systems (e.g., GRS 1915+105: Greiner, Morgan,
\& Remillard 1996; Belloni et al. 1997; Taam, Chen, \& Swank 1997; GS
2023+338: Terada et al. 1994; Oosterbroek et al. 1997). But a major
difference is that the intrinsic luminosities of those sources were
much higher (about $> 10^{39}$ erg \pers; see, e.g., Belloni et
al. 1997; Terada et al. 1994) than for V4641 Sgr.  Another unusual
property of V4641 Sgr is the very short time span of its bright X-ray
event ($< 10$ hours). In this respect, V4641 Sgr is similar to the
recently discovered transient CI Cam (Smith et al. 1998). However,
several differences are also present. The outburst of CI Cam was on
slightly longer time scales (days) and much more smooth (Belloni et
al. 1999) than the event observed for V4641 Sgr, which exhibited very
strong variability.  So, although V4641 Sgr resembles several sources
in some of its behavior, it differs from each of them significantly in
other respects.

Reanalysis of the ASM archive revealed several similar short lived
events for V4641 Sgr as the Sept. 15 event (Smith et al. 1999c), which
went previously unnoticed.  The Sept. 15 event was directly noticed
because ({\it a}) more attention to V4641 Sgr was paid because of its
sudden increase in the optical (Stubbings 1999) and ({\it b}) this
event was brighter (2--30 times) than the others.  The short life
times of the events and the strong flux fluctuations indicate that the
accretion is very unstable and highly irregular and not much accretion
takes place.  The fact that several events went unnoticed suggest that
many sources with similar events also fail to get noticed, especially
when they are at a greater distance than V4641 Sgr and have therefore
lower fluxes.  The exact number of such sources in our galaxy is
difficult to estimate because of the uncertainties in the distances
and the recurrence time scales of their outbursts.  However, if V4641
Sgr harbors a black hole, it is clear that a sizeable number of the
black holes in our galaxy could be present in V4641 Sgr like systems.

\clearpage

\clearpage

\begin{references}

\reference{}Ayani, K. \& Peiris, T.C.
1999,
\iaucirc 7254


\reference{}Barret, D. \& Vedrenne, G.
1994,
\apjs, 92, 505

\reference{}Belloni, T., M\'endez, M., King, A. R., van der Klis, M., 
\& van Paradijs,
1997,
\apj, 479, L145

\reference{}Belloni, T., Dieters, S., van den Ancker, M. E., Fender,
R. P., Fox, D. W., Harmon, B. A., van der Klis, M., Kommers, J. M.,
Lewin, W. H. G., \& van Paradijs, J.
1999,
\apj, in press (astro-ph/9907124)

\reference{}Charles, P. A., Shahbaz, T.,  Geballe, T.
1999
\iaucirc, 7267

\reference{}Djorgovski, S. G., Gal, R. R., Mahabal, A., Galama, T.,
Bloom, J., Rutledge, R., Kulkarni, S., \& Harrison, F.
1999,
ATEL $\#$44


\reference{}Ford, E. C., van der Klis, M., M\'endez, M., van Paradijs,
J., \& Kaaret, P.
1999,
\apj, 512, L31



\reference{}Gaensler, B. M., Campbell-Wilson, D., Hunstead, R. W., \&
Sault, R. J.  1999, 
\iaucirc 7256


\reference{}Greiner, J., Morgan, E. H., \& Remillard, R. A.
1996,
\apj, 473, L107




\reference{}Homan, J., Wijnands, R., \& van der Klis, M.
1999, \iaucirc, 7121


\reference{}Hjellming, R. M., Rupen, M. P., \& Mioduszewski, A. J.
1999a,
\iaucirc, 7254

\reference{}Hjellming, R. M., Rupen, M. P., \& Mioduszewski, A. J.
1999b,
\iaucirc, 7265

\reference{}Hjellming, R. M., et al.
1999c,
\apj, letters in prep.


\reference{}in 't Zand, J., Heise, J., Bazzano, A., Cocchi, M., Di
Ciolo, L., \& Muller, J. M. 
1999,
\iaucirc, 7119

\reference{}Liller, W.
1999,
\iaucirc, 7254


\reference{}Markwardt, C. B., Swank, J. H., \& Marshall, F. E.
1999a,
\iaucirc, 7120

\reference{}Markwardt, C. B., Swank, J. H., \& Morgan, E. H.
1999b,
\iaucirc 7257




\reference{}Nowak, M. A., Vaughan, B. A., Wilms, J., Dove, J. B., \&
Begelman, M.
1999a,
\apj, 510, 874

\reference{}Nowak, M. A., Wilms, J., \& Dove, J. B.
1999b, 
\apj, 517, 355

\reference{}Oosterbroek, T., van der Klis, M., van Paradijs, J.,
Vaughan, B., Rutledge, R., Lewin, W. H. G., Tanaka, Y., Nagase, F.,
Dotani, T., Mitsuda, K., \& Miyamoto, S.
1997,
\aap, 321, 776

\reference{}Remillard, R. A., McClintock, J. E., Sobczak, G. J.,
Bailyn, C. D., Orosz, J. A., Morgan, E. H., \& Levine, A. M.
1999, \apj, 517, L127



\reference{}Rutledge, R. E., Bildsten, L., Brown, E. F., Pavlov, G., \&
Vyatcheslav E. Zavlin
1999
\apj, in press (astro-ph/9909319)


\reference{}Smith, D., Remillard, R., Swank, J., Takeshima, T., \& Smith,
E. 
1998,
\iaucirc, 6855

\reference{}Smith, D. A., Levine, A. M., \& Morgan, E. H.
1999a,
\iaucirc, 7253

\reference{}Smith, D. A., Levine, A. M., \& Morgan, E. H.
1999b,
ATEL $\#$43

\reference{}Smith, D. A., et al.
1999c,
\apj, Letters in preparation


\reference{}Stubbings, R
1999,
\iaucirc, 7253

\reference{}Taam, R. E., Chen, X., \& Swank, J. H.
1997,
\apj, 485, L83

\reference{}Terada, K., Miyamoto, S., Kitamoto, S., \& Egoshi, W.
1994,
\pasj, 46, 677


\reference{}van der Klis, M.
1994,
\apjs, 92, 511


\reference{}van der Klis, M.  1995, In: {\it X-ray Binaries},
W. H. G. Lewin, J. van Paradijs, \& E. P. J. van den Heuvel (eds.),
Cambridge University Press, p. 252


\reference{}van der Klis, M., 1998, in The Many Faces of Neutron
Stars, ed. R. Buccheri, J. van Paradijs, \& M. A. Alpar (NATO ASI
Ser. C; Dordrecht: Kluwer), 337

\reference{}van der Klis, M. 1999, in the proceedings of the Third
William Fairbank Meeting, in press (astro-ph/9812395)

\reference{}van der Klis, M., Hasinger, G., Stella, L, Langmeier, A.,
van Paradijs, J., \& Lewin, W. H. G.
1987, \apj, 319, L13


\reference{}Wijnands, R. \& van der Klis, M. 
1999,
\apj, 514, 939


\end{references}
\end{document}